# Quantitative assessment of fitting errors associated with streak camera noise in Thomson scattering data analysis

G. F. Swadling[1,a)], C. Bruulsema[2], W. Rozmus[2], J. Katz[3]

[1]Lawrence Livermore National Laboratory, 7000 East Av., Livermore, CA, 94550, USA
[2]Department of Physics, University of Alberta, Edmonton, Alberta, Canada, T6G 2E1
[3]Laboratory for Laser Energetics, University of Rochester, Rochester, NY 14623, USA

[a)] Corresponding author: swadling1@llnl.gov

Thomson scattering measurements in High Energy Density experiments are often recorded using optical streak cameras. In the low-signal regime, noise introduced by the streak camera can become an important and sometimes the dominant source of measurement uncertainty. In this paper we present a formal method of accounting for the presence of streak camera noise in our measurements. We present a phenomenological description of the noise generation mechanisms and present a statistical model that may be used to construct the covariance matrix associated with a given measurement. This model is benchmarked against simulations of streak camera images. We demonstrate how this covariance may then be used to weight fitting of the data and provide quantitative assessments of the uncertainty in the fitting parameters determined by the best fit to the data and build confidence in the ability to make statistically significant measurements in the low signal regime, where spatial correlations in the noise become apparent. These methods will have general applicability to other measurements made using optical streak cameras.

## I. INTRODUCTION

Thomson scattering (TS) is a powerful measurement technique, capable of providing experimenters with diagnostic access to many of the key parameters that characterize plasma physics experiments. Fitting the shape of the scattered spectrum can provide measurements of electron density $n_e$, electron temperature $T_e$, flow velocity $\vec{v}$, ion temperature $T_i$, ionization state $\bar{Z}$, electron drift velocity $\vec{v}_e$ [1]. Over the last few decades this technique has been developed and applied with ever-increasing finesse to probe the dynamics of High Energy Density (HED) and Inertial Confinement Fusion (ICF) experiments conducted at large scale experimental facilities [2–12], providing new insights into the physics of plasmas under extreme conditions.

Thomson scattering measurements are typically made by probing the plasma using a tightly focused, monochromatic laser beam and recording the spectrum of light scattered out of this probe beam by the plasma using a grating spectrometer coupled to a detector, typically a streak camera or gated imager.

To extract measurements of the underlying plasma parameters from the spectrum of the scattered light, the data must be fitted using a theoretical model describing the amplitude spectrum of thermally excited electron density fluctuations. Reliable measurements of the plasma parameters therefore require accurate theoretical models to describe the amplitudes of the plasma fluctuations and the instrumental effects that are imprinted on the measured spectra. Robust methods for fitting the data using these models are required. Furthermore, if we wish to make strong scientific statements regarding the data then it is imperative that these fitting methods are capable of producing reliable quantitative assessments of the uncertainty in our measurements.

Understanding the signal requirements to make a measurement with a given precision is also important. The maximum probe laser intensity that can be used to measure the parameters of a given plasma are limited by the potential perturbative effects of the probe beam on the plasma. If the probe intensity is too high then the probe can induce significant plasma heating, disturbing and therefore invalidating the measurement [14]. A fundamental limit on usable probe intensity come when the pondermotive pressure of the probe beam exceeds the plasma pressure,

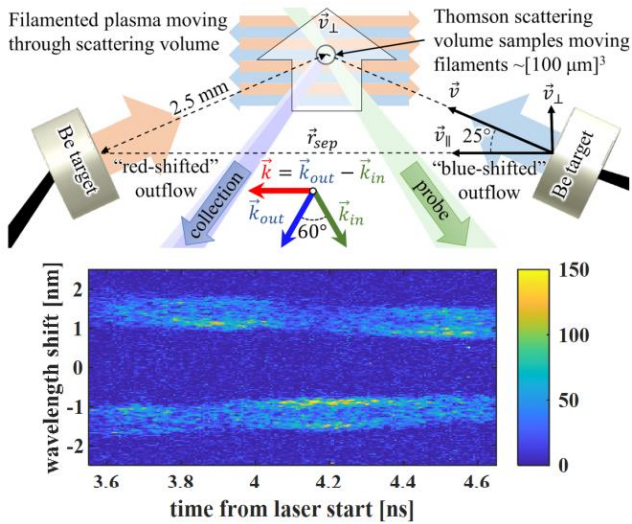

FIG. 1 a) Diagram of the experimental setup used in the experiment, including the Thomson scattering geometry. b) Example of the Thomson scattering data collected in the experiment. This is the ion-acoustic wave feature of the spectrum. The data in this plot has undergone a fluence-compensated warp correction to linearize the axes in both time and wavelength, similar to that described in ref [13]. Reproduced with permission from PRL. 124, 215001 (2020) Copyright 2020 American Physical Society [8].

leading to filamentation and spraying of the probe beam. While signal can be increased by increasing the size of the scattering volume, for example by using a phase plate to increase the size of the probe beam focal spot, this will results in an accompanying loss of measurement localization which must be accounted for in the analysis of the data [9]. Increases to the collection aperture (f-number) can also be used to increase signal, but this can become challenging optically beyond a certain point and can lead to signal blurring due the increasing range of scattering vectors that are sampled. These constraints motivate a desire on the part of the experimenter to collect Thomson scattering data at the minimum acceptable probe power sufficient to meet the measurement uncertainty requirements.

In this paper we present analysis of the effects of streak camera detector noise on measured Thomson scattering data and demonstrate a method of accounting for this noise when fitting the data with a theoretical scattering model to assess the resulting errors in the best fits to the data. This method is illustrated through the example of a set of Thomson scattering data collected in experiments conducted at the OMEGA laser facility at the university of Rochester's Laboratory for Laser Energetics (LLE). The results of these experiments have recently been published [8]. The data collected in these experiments were used to infer the strength of magnetic fields driven by the ion Weibel instability in interpenetrating HED plasmas [11]. The data were characterized by a low signal to noise which complicated accurate fitting of the data and estimation of the errors in the fitted parameters. In particular, the measurement of the plasma current was found to be very sensitive to the amplitude modulations introduced by the streak camera noise. In this paper we present a more rigorous review of the fitting technique and of the methods used to determine the uncertainties in the best-fit parameters. We benchmark this method using a Monte-Carlo method to simulate many noisy measurements of the same underlying spectrum. These methods will find application not only in the analysis of Thomson scattering data taken in many other HED experiments but also in any experiments that use optical streak cameras to record data.

We note that the scope of this paper addresses only the errors introduced by detector noise. There are many other potential sources of measurement error that must be considered in order to properly determine confidence in our measurements, such as errors in calibration, variations in instrument sensitivity with wavelength, the presence of background radiation and errors associated with using insufficient physical models to fit the data. Many of these topics have been treated elsewhere and are therefore not treated in detail here, although we do briefly discuss how the methods described here may be extended to cover other sources of error.

The remainder of this paper is structured as follows: In section II we provide a summary of the experimental setup and results reported in [8]. In section III we present a model of measurement noise introduced by optical streak cameras.

In section IV we discuss the method of fitting experimental data, weighted using a covariance model for the streak camera noise developed in section III. This section also shows how this streak camera noise model can be used to make quantitative assessments of the error in the resulting best fits to the data. Finally in section V we provide a worked example, first measuring the noise parameters of the streak camera, then fitting the experimental data presented in FIG. 1.

## II. SUMMARY OF EXPERIMENTAL DATA

The campaign of experiments that motivated this work focused on an investigation of the development of the ion-Weibel instability between pairs of interpenetrating plasmas streams. These experiments and the analysis and interpretation of these data are discussed in detail in our recent publication [8]. The plasma streams studied in these experiments are produced via direct laser heating of the surfaces of a pair of planar beryllium foils. Plumes of plasma expand from the foils with peak velocities of 1500 kms$^{-1}$, and at electron densities of ~$10^{19}$ cm$^{-3}$. A diagram of the experimental setup and an example plot of the Thomson scattering data collected in this experiment are reproduced in FIG. 1. The data plot shows the ion acoustic wave feature of the Thomson spectrum. This feature reflects scattering from low-frequency ion-acoustic fluctuations in the overall mass density of the plasma. The spectrum contains two gross features, one red-shifted and one blue-shifted. These features exhibit an anti-corelated modulation in intensity. Each feature reflects scattering off one of the two counterpropagating, interpenetrating plasma flows. For the geometry used in our experiments, the wavevector $|\vec{k}|$ of the density fluctuations observed by the Thomson scattering diagnostic lies parallel to the vector separating the centers of the two targets, and therefore the scattered spectrum is specifically sensitive to the approach velocity component of the two streams ($\vec{r}_{Sep}$ in FIG. 1). The Doppler shift due to this flow velocity is $\delta\omega = \vec{k} \cdot \vec{v}$. The equal but opposite velocities of the two interpenetrating flows in the direction $\hat{k}$ give rise to equal but opposite Doppler shifts, allowing the corresponding spectral features to be separately observed. The temporal modulation in the intensities of the two features reflects underlying modulations in the densities of the two streams within the Thomson scattering volume due to the development of stream filamentation. In the paper we demonstrated that analysis of this feature can provide measurements not only of the scale size and density contrast of these filaments, but also of the underlying current density modulation, allowing inference of the magnetic field strength in the plasma. This measurement required fitting to subtle variations in the relative amplitude of the ion acoustic peaks, caused by asymmetric Landau damping arising due to the shifting of the center of the electron distribution function associated with the plasma current. This measurement is particularly sensitive to the measurement noise introduced by the streak camera.

## III. FITTING OF THOMSON SCATTERING DATA IN THE PRESENCE OF NOISE INTRODUCED BY OPTICAL STREAK CAMERAS

The Thomson scattering diagnostic implemented on the OMEGA laser system at LLE uses a pair of ROSS [15] optical streak cameras to record data when operated in their time-resolved mode. Similar cameras are also used on the new 5th Harmonic optical Thomson scattering system currently being commissioned at the National Ignition Facility (NIF) and therefore the analysis presented here will have relevance to data taken on that system as well. The raw Thomson scattering data presented in FIG. 1 has a characteristic "speckled" texture. This texture reflects the measurement noise introduced by the streak camera. The noise is not spatially independent across the image but is instead correlated over small spatial scales. A proper understanding of the origin of this noise is required to account for it and to access its impact on the achievable accuracy of Thomson scattering measurements.

A schematic diagram of an optical streak tube is provided in FIG. 2 for reference. The streak camera measurement process can be broken down into a series of steps. First, scattered light is collected from the experiment by a telescope, dispersed by a spectrometer and transported to the photocathode of the camera's vacuum tube. Secondly, at the photocathode this optical signal is converted to an electron beam signal; bound electrons are freed from the photocathode by incoming photons via the photoelectric effect and accelerated through the streak tube by an extraction electric field. The accelerated electrons are focused by electro-static lenses and then deflected ("swept") by a time varying electric field before finally striking a phosphor screen. Sweep durations used for Thomson scattering measurements typically vary between 5 – 30 ns for long-pulse HED experiments. At the phosphor, the energetic electrons produce a pulse of light which is transported via a fiber-optic image coupler to a CCD image sensor.

The limited optical power delivered to the photocathode introduces Poisson noise to the measurement due to photon quantization. The limited quantum efficiency of the photocathode further amplifies this effect; the number of streak-tube electrons sets the true statistical noise floor of the measurement. At the phosphor screen the pulse of light produced by each photoelectron produces many CCD counts which are distributed over a small range of adjacent pixels on the CCD, leading to the spatially correlated measurement noise that is observed in the recorded data (as seen in the "texture" observed FIG. 1 b)). Finally, additional noise is introduced by the CCD itself, which has a characteristic read noise associated with thermal fluctuations in the counting electronics.

In this section we present an analytical model which statistically approximates the streak camera noise generation mechanism. This model is benchmarked against a numerical model of the streak camera noise that uses random number generators to simulate the underlying noise generation mechanisms. The resulting analytical expression for the expected measurement covariance matrix, $\vec{K}_{Noise}$, can be efficiently used both to weight fittings to experimental data using non-linear regression techniques, and to quantitatively assess the uncertainty in the resulting best-fit parameters.

### A. Modeling streak camera noise using randomly generated numbers.

The noise introduced by the streak camera can be numerically simulated using a random number generator. FIG. 3 shows the step-by-step calculation of a single example of a simulated TS spectrum measurement, using a random number generator to account for the effect of Poisson noise. This section describes how these images are generated.

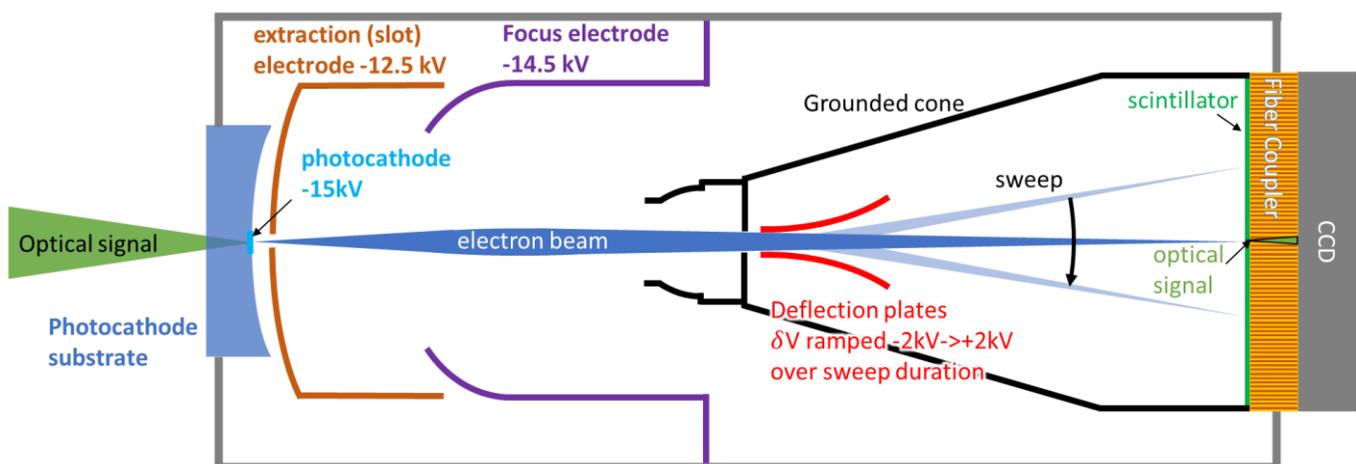

FIG. 2 Schematic diagram of an optical streak vacuum tube. The optical signal is focused at the photocathode where it liberates electrons via the photo-electric effect. Freed electrons are accelerated due to the electric field between the photocathode and the extraction electrode. The resulting beam is focused by the electric field structure produced by the focus electrodes, forming an image of the photocathode. This diagram shows a cross section of the tube taken in the plane defined by the tube axis and the temporal axis of the camera. The photocathode is an extended object (in and out of the page) which is imaged as a 1D line focus at the scintillator. The energetic electron strike the scintillator producing light which is coupled to the CCD image sensor via a fiber optic bundle.

### 1. *Plasma Model and scattering calculation.*

FIG. 3 a) shows plots of the time history of the plasma parameters which were used to generate the synthetic spectra. These plasma parameters represent a pair of interpenetrating ion streams, similar to those described in our recent publications [8,11]. These parameters are used as an example so that the results of the simulation can be directly compared with measured streak camera data provided in FIG. 1. The plasma is modeled as a pair of two counterpropagating, interpenetrating Beryllium ion populations and a single electron population. Distribution functions are assumed to be Maxwellian. The temperatures are assigned as constant $T_e = 500$ eV and $T_i = 600$ eV. Sinusoidal modulations in the relative ion densities fractions of the two streams $n_j/\sum n_j$ and the overall plasmas current density $\vec{J}$, are imposed and flow velocities of the two streams steadily decrease, consistent with the evolution of the plasma parameters inferred from the experimental data [8].

The spectral flux $\Phi_{e,\lambda}$ (W nm$^{-1}$) that should be collected by the detector, due to scattering from this plasma is calculated using a model for the scattered power [1]:

$$\Phi_{e,\lambda}(\lambda_{out}) = \frac{\bar{E}_e n_e r_e^2 c V_S \Omega_C}{\lambda_{out}^2}\left(\frac{2\lambda_{in}}{\lambda_{out}} - 1\right) S(\vec{k}, \omega) \quad (1)$$

Where $\bar{E}_e$ is the average flux density of the probe beam (power per unit area), $\lambda_{in}$ is the probe wavelength, $\lambda_{out}$ is the scattered wavelength, $\Omega_C$ is the solid angle collected by the diagnostic, $V_S$ is the scattering volume and $r_e$ is the classical electron radius. $S(\vec{k}, \omega)$ is the dynamic structure factor, describing the spectral density of thermally excited electron density fluctuations of wavevector $\vec{k}$ and frequency $\omega$.

$$S(\vec{k},\omega) = \frac{2\pi}{|\vec{k}|}\left|1 - \frac{\chi_e}{\epsilon}\right|^2 f_{eo} + \frac{2\pi}{|\vec{k}|}\sum_j \frac{\bar{Z}_j^2 n_j}{n_e}\left|\frac{\chi_e}{\epsilon}\right|^2 f_{jo} \quad (2)$$

$$\vec{k} = \vec{k}_{out} - \vec{k}_{in} \quad (3)$$

$$|\vec{k}_{in}| = \frac{\sqrt{\omega_{in}^2 - \omega_{pe}^2}}{c}; |\vec{k}_{out}| = \frac{\sqrt{\omega_{out}^2 - \omega_{pe}^2}}{c} \quad (4)$$

$$\omega = \omega_{out} - \omega_{in} \quad (5)$$

$$\omega_{in} = \frac{2\pi c}{\lambda_{in}}; \omega_{out} = \frac{2\pi c}{\lambda_{out}} \quad (6)$$

The scattering geometry used in these calculations matches that used in the experiment, as illustrated in FIG. 1 a). In these and the following equations, subscript $e$ is used to label the parameters of the electron population, $j$ the two ion populations and subscript $s$ is used to indicate that the equation may be used for either type of population. Parameters $n_j$ and $\bar{Z}_j$ are the density and charge state of each ion species making up the plasmas, and $f_{e0}$, $f_{j0}$ are the velocity distributions of each plasma species. For arbitrary distribution functions the electric susceptibilities $\chi_s$ and the total permittivity $\epsilon$ are calculated:

$$\chi_s(\omega,\vec{k}) = \frac{\omega_{ps}^2}{|\vec{k}|^2}\int_{-\infty}^{\infty}\frac{1}{\omega/|\vec{k}| - v}\frac{df_{s0}(v)}{dv}dv$$

$$\varepsilon = 1 + \sum_s \chi_s \quad (7)$$

Where $\omega_{ps}$ is the species-specific plasma frequency.

$$\omega_{ps} = \sqrt{\frac{n_s e^2 \bar{Z}_s}{\epsilon_0 m_s}} \quad (8)$$

For the scope of this paper, we make the simplifying assumption that the distribution functions are Maxwellian.

$$f_{s0} = \sqrt{\frac{1}{\pi \vec{v}_{Ts}^2}}\exp\left(-\frac{|\vec{v}_{\phi s}|^2}{v_{Ts}^2}\right) \quad (9)$$

$$v_{Ts} = \sqrt{\frac{2k_B T_s}{m_s}}; \quad \vec{v}_{\phi s} = \frac{\omega - \vec{k}\cdot\vec{v}_s}{|\vec{k}|} \quad (10)$$

Here $\vec{v}_{Ts}$ is the thermal velocity and $\vec{v}_{\phi s}$ is the species-specific Doppler-corrected phase velocity, which accounts for the bulk flow velocity $\vec{v}_s$ of the species with respect to the reference frame in which the scattering is measured. The mean drift velocity of the electron population $\vec{v}_e$ is calculated based on the weighted mean $\vec{v}_j$ and the plasma current density $\vec{J}$:

$$\vec{v}_e = \frac{1}{n_e}\sum_j n_j \bar{Z}_j v_j - \frac{J}{en_e} \quad (11)$$

For Maxwellian distributions the susceptibilities can be reduced to the following expression:

$$\chi_s = -\alpha_s^2 Z'\left(\frac{\vec{v}_{\phi s}}{v_{Ts}}\right) \quad (12)$$

Where $Z'$ is the derivative of the plasma dispersion function, $\alpha_s$ is the species-specific, dimensionless scattering parameter and $\lambda_{Ds}$ is the species-specific Debye length.

$$Z'(\zeta) = \frac{1}{2\sqrt{\pi}}\int_{-\infty}^{\infty}\frac{e^{-t^2}}{(t-\zeta)^2}dt \quad (13)$$

$$\alpha_s = \frac{1}{|\vec{k}|\lambda_{Ds}}; \lambda_{Ds} = \sqrt{\frac{\epsilon_o k_B T_s}{n_s \bar{Z}_s^2 e^2}} \quad (14)$$

These equations are used to numerically calculate the time dependent spectral flux collected by the Thomson scattering instrument [16]. The next step is to account for the chain of optical components that transport the collected light to the photocathode. This optical path will have a characteristic transport efficiency $\epsilon_\lambda$, the product of the many individual reflectivities and transmittivities of the individual optics. The wavelength dependence of the efficiency can often be neglected for ion acoustic wave (IAW) measurements due to the small ratio between the measured bandwidth and the typical wavelength scale over which the efficiency changes. For the purposes of this paper, we have used a single wavelength–independent efficiency as a free parameter to tune the amplitude of our simulated data to the amplitude of our measured data. For broadband electron plasma wave (EPW) measurements the wavelength dependence of this parameter is very important, introducing gross distortions to the measured spectral shape of the scattered light and its wavelength-dependence therefore cannot be neglected.

The effect of the spectrometer is to disperse the transported light across the photocathode resulting in a bandwidth per recorded data pixel of $\delta\lambda_{px}$, while the streak camera disperses the flux in time, with a sweep rate that can be characterized by a dwell time per pixel $\delta t_{px}$. The radiant energy delivered by the diagnostic to the photocathode per CCD pixel is then simply calculated:

$$Q_e = \Phi_{e,\lambda}\, \epsilon_\lambda\, \delta\lambda_{px}\, \delta t_{px} \quad (15)$$

For our example data $\delta\lambda_{px} \approx 9 \times 10^{-3}$ nm and $\delta t_{px} \sim 5$ ps. The resulting $Q_e$ corresponding to our imposed variation of plasmas parameters is plotted in FIG. 3 b). This is the optical energy per pixel that would be expected to be delivered to the streak camera photocathode, assuming the instrument introduces no spectral or temporal broadening. For the purpose of this calculation, $\epsilon_\lambda$ was scaled to produce a $Q_e$ that would match the amplitude of the data presented in FIG. 1.

## 2. *Accounting for sources of broadening.*

The spectral and temporal performance of the Thomson scattering instrument are limited by a variety of mechanisms, with contributions coming from both the design of the spectrometer and from the performance characteristics of the streak camera. The effects of the various broadening mechanisms can be lumped together and modeled using an instrument envelope function which can be convoluted with theoretical spectra to model the shape of the data that will be measured by the instrument. For the calculations presented in this paper this envelope function is approximated using the following normalized filter kernel:

$$\mathbf{g} = \frac{1}{\sum_{i,j} g^*_{ij}}\left[g^*_{ij}\right] \quad (16)$$

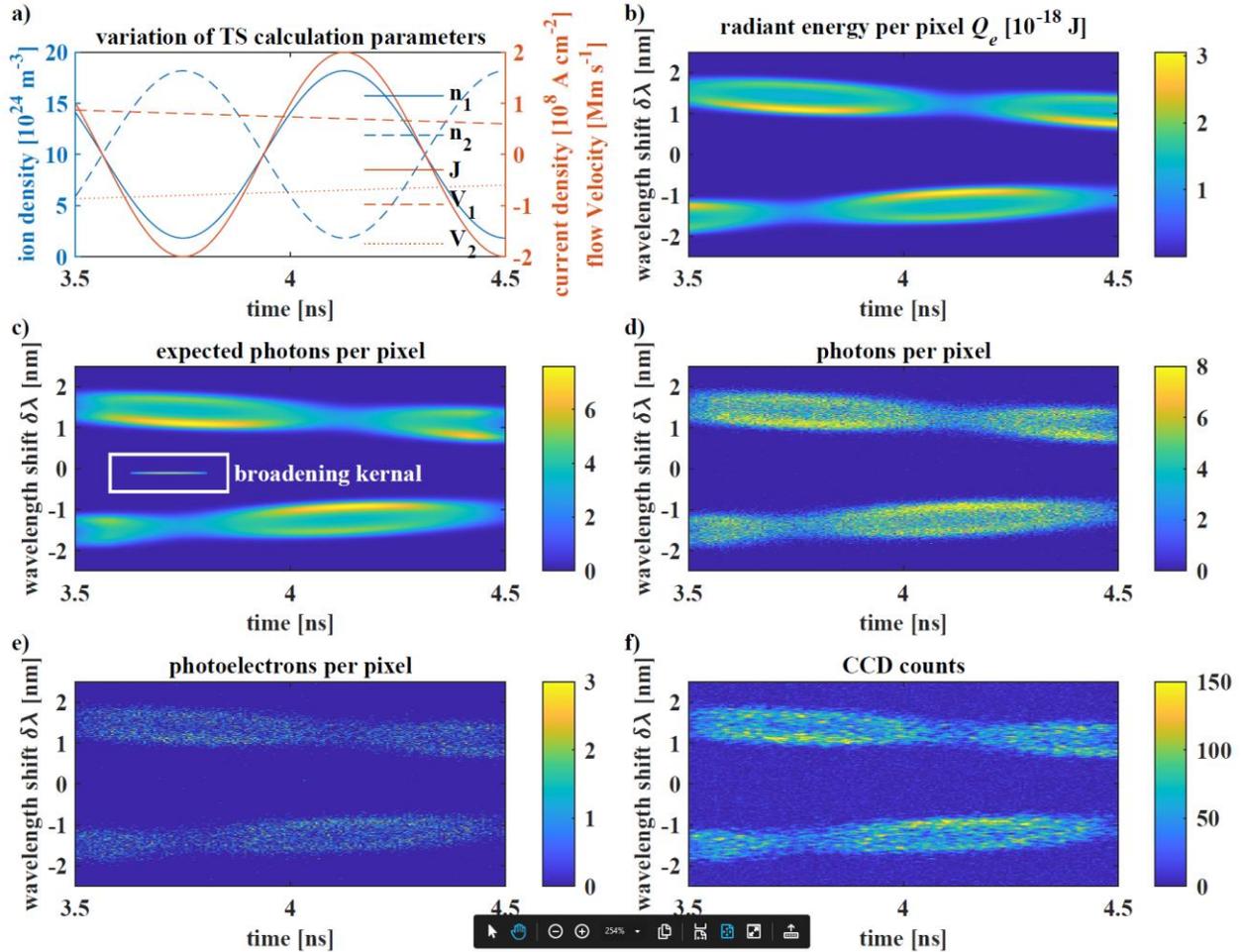

FIG. 3 Illustration of the steps in generating a synthetic streak image using a random number generator based on our model of streak camera noise. a) Imposed variation in plasma parameters. b) Synthetic spectrum calculated using equation (1). c) Effect of spectral and temporal broadening imposed by the instrument. d) Example of a photon distribution corresponding to expected photon density in (c). e) Further reduction in signal/noise due to conversion from photons to photo-electrons. The dynamic range of images d) and e) were adjusted to include 99% of all pixel values. Maximum values calculated for statistical outliers were 15 photons and 5 electrons respectively. f) Effects of optical coupling from phosphor to the CCD and CCD read noise. This example of a final recorded images compares well with the data recorded in the experiment (FIG. 1).

$$g_{ij}^* = \sqrt{\delta_t^2 - i^2} \exp\left(-\frac{j^2}{2\sigma_\lambda^2}\right)$$
$$i = \lfloor -\delta_t \rfloor : \lceil \delta_t \rceil$$
$$j = \lfloor -5\sigma_\lambda \rfloor : \lceil 5\sigma_\lambda \rceil$$

Where $\sigma_\lambda$ is the standard deviation of a gaussian broadening in wavelength and $\delta_t$ is the half width of a hemispherical broadening in time. Bracket notation $\lceil\ \rceil$ & $\lfloor\ \rfloor$ indicate rounding up or down to the nearest integer respectively.

The spectral standard deviation was determined using a calibration lamp to record the spectrum of narrow emission lines. For the IAW spectrometer configuration used in our experiment $\sigma_\lambda = 2.36$ px. The dispersion in this configuration is 0.0089 nm px$^{-1}$, so that this corresponds to 0.021 nm standard deviation in wavelength. For the IAW spectrometer the spectral resolution is chiefly limited by the range of input angles introduced by the 100µm dia. pinhole used as the input field stop, but the broadening associated with the electro-optical imaging performance of the streak tube is of a similar magnitude. Broadening due to the diffraction limited focusing performance of the spectrometer is negligible for the IAW system but may be important for other instruments.

The temporal broadening of the system is dominated by the pulse-front tilt introduced by the spectrometer design. This effectively smears the arrival time of light at the photocathode. The broadening has a hemispherical shape due to the circular shape of the signal beam impinging on the grating. For a grating spectrometer the magnitude of temporal smearing due to the pulse front tilt can be estimated [17]:

$$\delta_t \sim \frac{1}{2} \frac{m\lambda}{c} \frac{Gw}{\cos\theta} \quad (17)$$

Where $w$ is the illuminated width of the grating, $G$ is the grating line density and $\theta$ is the grating angle with respect to the m = 0, mirror like orientation. For the IAW spectrometer configuration used in our experiments $G = 1800$ mm$^{-1}$, $w = 50$ mm, $\theta \sim 25°$ and therefore $\delta t \sim 90$ ps, or ~16 px. The electro-optical resolution of the streak camera can also be measured using the calibration line source in static mode and yielded standard deviation of 1.9 px. This is negligible compared to the pulse front tilt and its effect has not included in equation (16).

The overall effect of this broadening on the measured spectrum is illustrated in FIG. 3 c), which shows the expected number of photons delivered to the photocathode per pixel:

$$\bar{N}_p = \frac{\lambda}{hc} (Q_e * \mathbf{g}) \quad (18)$$

An amplitude-scaled plot of $\mathbf{g}$ is inset within this image to provide a visual guide to the overall scale of this blurring effect.

### 3. *Modeling streak camera noise using a random number generator.*

As FIG. 3 c) illustrates, the radiant energy delivered to the streak photocathode per detector pixel is comparable to the photon energy and quantization effects must therefore be considered. The actual number of photons observed per pixel during any given measurement follows Poisson statistics. For an expected number of photons $\bar{N}_p$ the probability of actually observing $N_p$ photons is:

$$P(N_p) = \frac{\bar{N}_p^{N_p} e^{-N_p}}{N_p!} \quad (19)$$

This effect can be numerically simulated using a random number generator to generate photon counts following the Poisson distribution. Applying this approach to the broadened spectrum plotted in FIG. 3 c) results in the photon map plotted in FIG. 3 d).

Photo-electric conversion further reduces the dynamic range of the measurement, acting effectively as the quantum "bottle-neck" in the streak camera and dominating the overall noise of the measurement. Conversion is characterized by a wavelength-dependent quantum efficiency $\epsilon(\lambda)$ – the probability that an incident photon will produce a photoelectron that will make it through the streak tube. For optical streak cameras typically $\epsilon(\lambda) \sim 0.1 – 0.2$. Since the photo-electric conversion process is a purely probabilistic process, with each conversion from photon to electron being statistically independent from the others, the overall result can be modeled using binomial statistics, such that if the expected number of photoelectrons is,

$$\bar{N}_e = N_P \epsilon(\lambda) \quad (20)$$

, then the probability of producing $N_e$ photoelectrons given $N_p$ photons is,

$$P(N_e) = \frac{N_p!}{N_e!(N_p - N_e)!} \epsilon^{N_e}(1-\epsilon)^{N_p - N_e} \quad (21)$$

This step can again be simulated using a random number generator, this time generating numbers following the binomial probability distribution. For illustration FIG. 3 e) shows a map randomly generated single photoelectron events (SPEs) using the photon map plotted in FIG. 3 d) as its input. Comparison of these images illustrates the resulting amplification of the measurement noise. Since conversion to photoelectrons is a binomial process, the overall conversion from expected photons at the photocathode to photoelectrons can be calculated directly in a single step, using Poisson statistics:

$$P(N_e) = \frac{\bar{N}_e^{N_e} e^{-\bar{N}_e}}{N_e!} \quad (22)$$

At the phosphor each single photo-electron event produces a pulse of light which is the fiber-coupled to the CCD image sensor. Imperfections in the coupling of this light leads to further modifications to the appearance of the data. Multiple CCD counts are recorded for each photoelectron that reaches the phosphor and these counts are distributed over a small area of the CCD. The effects can be parameterized by a CCD gain $G$, characterizing the expected number of counts per photoelectron, by a point-spread standard deviation $\sigma_{px}$, describing the size of the 2D gaussian spatial distribution of these counts on the sensor and by a noise factor $F^2$, which quantifies the additional noise introduced by the amplification process above that expected due to the underlying counting statistics. For the optical streak cameras used in our experiments typical values are $G \sim 135$ and $\sigma_{px} \sim 1$ and $F^2 \sim 1.05$ (see section V.A for details) [18,19]. To model these effects in the measurement simulation, the expected number of CCD counts is calculated, and the required additional amplifier noise is added using a normally distributed random number generator scaled by the signal dependent standard deviation given below:

$$N_C = G\bar{N}_e = \epsilon \bar{N}_p G \quad (23)$$
$$\sigma_{N_C} = \sqrt{(F^2 - 1)(N_e G)} \quad (24)$$

This is then convoluted with a gaussian kernel to simulate the spreading of the signals across multiple pixels:

$$N_c = (N_p G) * \mathbf{g} \quad (25)$$
$$\mathbf{g} = [g_{ij}] = \frac{1}{2\pi\sigma_{px}^2} \exp\left(-\frac{i^2 + j^2}{2\sigma_{px}^2}\right) \quad (26)$$
$$i, j = \lfloor -|8\sigma_{px}|\rfloor : \lceil|8\sigma_{px}|\rceil$$

Finally, CCD read out noise is added. This noise is also simulated using a normally distributed random number generator, scaled by a standard deviation of $\sigma_{RN}$ 11 px and rounded. The randomly generated noise is simply added to the image. The overall effect of these CCD readout effects is illustrated in FIG. 3 f).

#### 4. *Determining the sample mean and sample covariance using the random image generation approach*

In an experiment we can only to make a single measurement of the Thomson scattering spectrum. Gathering data in sufficient volumes to perform a proper statistical analysis for HED experiments is often resource prohibitive. In experiments such as the one provided as an example in the paper, the dynamics are unstable and therefore the experiments are in any case not expected to be directly reproducible.

The Monte-Carlo approach outlined in the previous section provides a means to perform many simulated measurements of the same underlying dynamics, allowing both the sample mean and sample covariance of a specific measurement to be determined. This provides us with a powerful tool to assess the potential errors in our measurements introduced by detector noise.

To determine the sample mean and covariance, first a center time $t$ and binning period $\Delta t$ must be selected. An image is then calculated over the required time period and the signal is extracted by summing in the temporal direction to produce a profile varying in intensity as a function of wavelength. This process is repeated over many random calculations of the measured signal to build up a statistically significant number (i=1,..,n) of numerically simulated example measurements.

$$S_i(\lambda) = \int_{t-\frac{\Delta t}{2}}^{t+\frac{\Delta t}{2}} N_{c_i} \, dt \quad (27)$$

The sample mean of these measurements is just the mean of all of all they simulated measures signals:

$$\bar{S}(\lambda) = \frac{1}{n}\sum_{i=1}^{n} S_i \quad (28)$$

The sample covariance matrix is calculated based on the residuals of the many example profiles from this mean:

$$R_i = S_i - \bar{S} \quad (29)$$
$$\bar{K} = \frac{1}{n-1}\sum_{i}^{n} R_i R_i^\intercal \quad (30)$$

### B. Analytic estimate of the expected covariance for streak camera measurements based on the expected signal.

In section III.A we illustrated how the streak camera introduces spatially correlated errors in the measurements at each CCD pixel due to the amplification of underlying Poisson noise. Proper fitting of the data and assessment of the errors in the determination of the fitting parameters requires us to be able to model and account of this correlated noise. Data analysis is normally carried out by fitting the spectrum at various points in time across the streak image. Data is typically binned over some temporal range $\Delta t$, to improve the statistical significance of the measurement to an acceptable level. This corresponds to some pixel width $n$ in the temporal direction of the streak image. The measured signal $S$ that is fitted to is therefore the sum over an image width n:

$$S(\lambda, t) = \sum_{n} N_{c_i} \quad (31)$$

This measured signal corresponds to an integral of the underlying theoretical scattering signal.

$$\bar{S}(\lambda, t) = \int_{t-\frac{\Delta t}{2}}^{t+\frac{\Delta t}{2}} \bar{N}_p \epsilon G \, dt \quad (32)$$

Ignoring the spread of the signal at the CCD, the variance $\sigma_S^2$ in the expected signal, $S(\lambda, t)$, introduced by the streak camera measurement noise can be estimated:

$$\sigma_S^2 = \left(\frac{d\bar{S}}{dN_e}\sigma_{N_e}\right)^2 + \left(\frac{d\bar{S}}{dG}\sigma_G\right)^2 \quad (33)$$

Where the first term is the variance associated with conversion from expected photons to actual photoelectrons and the second term describes the additional variance introduced by the gain process that converts those photoelectron to a CCD signal. Conversion to photoelectrons in a Poisson process and the variance is therefore:

$$\sigma_{N_e}^2 = \frac{\bar{S}}{G} \quad (34)$$

The variance in the gain can be parametrized by a noise factor $F^2$, which describes the additional variance associated with gain process:

$$\sigma_G^2 = \frac{G^3(F^2 - 1)}{\bar{S}} \quad (35)$$

The overall expected variance for the signal is therefore:

$$\sigma_S^2 = G^2\frac{\bar{S}}{G} + \left(\frac{\bar{S}}{G}\right)^2 \frac{G^3(F^2 - 1)}{\bar{S}} \quad (36)$$
$$\sigma_S^2 = \bar{S}G F^2 \quad (37)$$

For optical streak cameras the noise factor has been measured and found to be typically $F^2 \sim 1.05 - 1.21$ [18].

To account for the grouping of CCD counts associated with each photo-electron detection event this signal variance must be converted into a covariance matrix $\mathbf{K_{Noise}}$:

$$\mathbf{K_{Noise}} = \left(\left(I\sigma_s^2\right) * \mathbf{g}\right) + \mathbf{I}(n\sigma_{RN}^2) \quad (38)$$

Here $\mathbf{I}$ is the identity matrix and the $*\mathbf{g}$ notation indicates convolution by a normalized 2D Gaussian kernel defined by the CCD spread standard deviation $\sigma_{px}$:

$$\mathbf{g} = [g_{ij}] = \frac{1}{2\pi\sigma_{px}^2} \exp\left(-\frac{i^2 + j^2}{2\sigma_{px}^2}\right) \quad (39)$$

$$i, j = \lfloor -|5\sigma_{px}| \rfloor : \lceil |5\sigma_{px}| \rceil$$

Equation (38) consists of two terms; the first encodes the spatial correlation of the covariance associated with the streak camera Poisson noise. This term is dependent on the expected signal $\bar{S}$. The second term accounts for the uncorrelated, uniform CCD read noise with standard deviation $\sigma_{RN}$, and is signal independent. This equation provides a means of calculating the expected covariance of any proposed fit to the data so that the quality of the fit may be properly weighted. The covariance can be updated through the iterative fitting process as the best fit is refined.

This analytic covariance model was benchmarked against the streak camera image simulation model outlined in section III.A. A set of 10000 random simulated images similar to FIG. 3 f) were generated. Each image was summed over a 0.1 ns period, centered at 3.9 ns to extract a spectral profile. The sample mean and standard deviation of this set of profiles was calculated using equations (28)(28 and (30). For the analytical calculations the expected signal $\bar{S}$ was found using equation (32), summing over the same 0.1 ns period using the data plotted in FIG. 3 c). An additional gaussian broadening $\sigma_{px}$ was applied to account for the effect of the photo-electron pixel spread. The expected covariance $\mathbf{K}$ was then calculated from this $\bar{S}$ using equation (38).

The results of this comparison are plotted in FIG. 4; in a) data from a single example simulated measurement is plot against the sample mean and the expected mean signal. As expected, the sample mean is in excellent agreement with the expected mean. In b) the sample and expected covariance are compared. These data are plotted on the same color scale and again show excellent agreement; c) compares the square root of the diagonal elements of these two covariance matrices, illustrating the agreement in amplitude. The naïve standard deviation is also plotted here. This is the expected standard deviation $X$ of the signal ignoring the correlation of errors introduced by conversion from photoelectrons to CCD counts, i.e. it assumes that all the noise is spatially independent.

$$X = \sqrt{\bar{S}GF^2 + n\sigma_{RN}^2} \quad (40)$$

## IV. FITTING EXPERIMENTAL DATA USING THE NEWTON-GAUSS METHOD WEIGHTED BY THE EXPECTED COVARIANCE

In section III.B we developed a model (Eqn (38)) that allows us to estimate the expected covariance matrix $\mathbf{K}$ directly from the expected signal. In this section we describe how this $\mathbf{K}$ can be used to weight the quasi-Newton non-linear least squares (NLLS) regression fitting method. Weighting the fit using $\mathbf{K}$ properly accounts for the expected correlations in the errors in the measured data that arise due to the noise introduced by the conversion from photoelectrons to CCD counts, effectively discounting errors that are spatially grouped over small spatial scales with respect to random errors. Once a best fit is found $\mathbf{K}$ may be used to estimate the errors in the determination of the best fit.

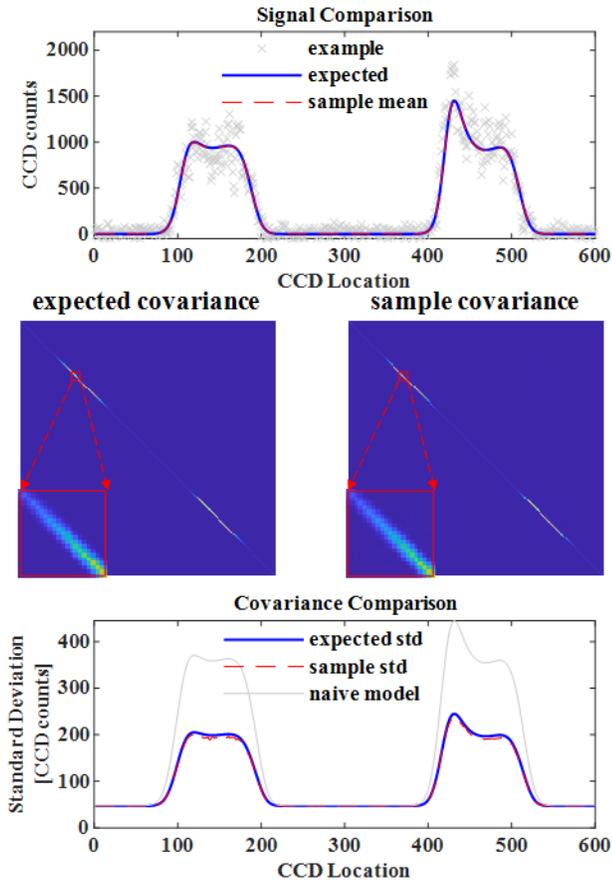

FIG. 4 Verification of the model for the expected standard deviation. a) plot of data, comparing expected mean signal in the absence of noise and the sample mean for 10000 calculations of the spectrum with randomly generated noise. Profiles are binned over 0.1 ns, centered at 3.9ns w.r.t plots in FIG. 3. One of these sample random "measurements" is provided for reference. b) Comparison of the sample covariance from the random calculations and the expected covariance calculated using equation (38), plotted on the same color scale, illustrating their similarity. Sub regions are expanded to illustrate the width introduced by the spread of CCD counts associated with each photoelectron c) Comparison of the sample and expected standard deviation (the square root of the diagonals of the covariance).

## A. The quasi-Newton fitting Method

A "quasi-Newton" gradient decent non-linear regression method [20,21] is used to fit the measured data to the theoretical model. We describe the general approach below. The measured data consists of $n$ measurements of some observable $y$ over a range of some variable $x$:

$$y(x) = y_1(x_1), \ldots y_n(x_n) \quad (41)$$

The observable is modeled using a function $f(x, \beta)$, a function variable $x$ and a set of $m$ fitting parameters $\beta = (\beta_1, \ldots, \beta_m)$. The residual function $r(\beta) = (r_1(\beta), \ldots, r_n(\beta))$ is defined as the difference between the data and fitting function $f(x, \beta)$ for a specific set of fitting parameters $\beta$.

$$r(\beta) = y - f(x, \beta) \quad (42)$$

The process of fitting seeks to minimize the covariance-normalized sum of the squares of the residuals,

$$S = (r^\intercal K^{-1} r) \quad (43)$$

Where $K$ is the covariance matrix describing the uncertainties in the measurement. The condition $S(\beta) = N$, where $N = n - m$ is the number of degrees of freedom in the fit (observations – free parameters), corresponds to the case where the distribution of the measurement errors is equal to that described by covariance matrix $K$. This condition would indicate a good fit to the data given our understanding of the measurement errors and can be used to estimate the range of acceptable fitting parameters (i.e. the error in the best fit).

Starting at some initial guess of the fitting parameters $\beta^0$, the Newton-Gauss algorithm proceeds by iteration:

$$\beta^{s+1} = \beta^s - H^{-1} g \quad (44)$$

Each iteration step calculates a correction to the current value of the best fits for fitting parameters $\beta^s$. The correction, $H^{-1} g$, is assessed at the current best fit $\beta^s$. Vector $g$ is the normalized gradient vector of $S$,

$$g = \left[\frac{\partial S}{\partial \beta_i}\right] = (r^\intercal K^{-1} J) \quad (45)$$

, matrix $H$ is an approximation of the normalized Hessian matrix of $S(\beta)$,

$$H = \left[\frac{\partial^2 S(\beta)}{\partial \beta_i \partial \beta_j}\right] \sim J^\intercal K^{-1} J + \cdots \quad (46)$$

, matrix $J$ is the Jacobian matrix of the residuals,

$$J = [J_{ij}] = \left[\frac{\partial r_i}{\partial \beta_j}\right] \quad (47)$$

and matrix $K$ is expected covariance of the measurement. The residuals $r$ are straightforward to calculate. The Jacobian $J$ matrix can be approximated using a finite difference method:

$$\frac{\partial r_i}{\partial \beta_j} \sim \frac{r_i(\beta + \vec{e}_j \delta_{\beta_j}) - r_i(\beta - \vec{e}_j \delta_{\beta_j})}{2 \delta_{\beta_j}} \quad (48)$$

Where $\delta_{\beta_j}$ are small steps in $\beta_j$ and $\vec{e}_j$ are the standard basis vectors. The expected covariance matrix $K$ must be estimated based on a model for the expected measurement noise, such as the one presented in this paper. In our case $K$ is a function of the fit $f(x, \beta)$ and is therefore recalculated on each iteration. Iteration of equation (44) is terminated once the required precision in the best fit for $\beta$ is met, i.e. when further iterative corrections to $\beta$ are smaller than some pre-specified precision. A precision of 0.1% was used for fits presented in this paper, which is at least an order of magnitude smaller than the errors associated with the determination of the best fits for any of the plasma parameters.

It should be noted that like any non-linear regression scheme, the Newton-Gauss method is susceptible to converging on local minima in $S$. To check that we have found the global minimum and therefore the actual best fit, the algorithm can be initiated from multiple starting conditions to find the solution with the lowest $S$.

## B. Modifications on the quasi-Newton algorithm

The method described in section IV.A assumes that the fit function can be locally approximated as quadratic in the region close to the fit in order to estimate the step size in $\beta$ required to reach the best fit. The direction in $\beta$ space of $H^{-1} g$ is the local direction of steepest decent in $S$ at the current values of $\beta$ and its amplitude is the expected correction step required to reach the minimum in $S$. In many cases this approximation is inadequate, producing steps that are too large and which can trigger a convergence failure of the fitting algorithm. This is particularly an issue if the algorithm introduces a step which moves some element of $\beta$ into a region of parameter space where the residual becomes insensitive to its variation.

This convergence failure can be mitigated by introducing some additional steps in the fitting algorithm. The iterative equation given in (44) can be modified with the addition of scalar factor $C_i$.

$$\beta^{s+1} = \beta^s - C_i H^{-1} g \quad (49)$$

The literature describes many different approaches to estimating $C_i$ [20,21]. For the work presented in this paper, a simple algorithm was used to reduce $C_i$ until $S$ was observed to decrease. This was sufficient for the current problem.

$$C_i = \frac{1}{4^{(i-1)}}; i = 1, 2, \ldots \quad (50)$$

Fitting proceeds by calculating $H^{-1} g$, applying the expected correction and calculating the new value of $S$. This value is compared to existing value of S. If $S$ decreases the algorithm proceeds as normal. If $S$ increases, then this indicates that the estimated step size was too large and the algorithm instead iterates in $i$, so that the amplitude of the correction is reduced by a factor of four. The $S$ is again calculated and compared to the value at $\beta^s$. This process continues until either the new value of $S$ is lower than the previous value, or the step size for the correction becomes smaller that the required accuracy of the best fit. Since $C_i$ does not modify the direction of $H^{-1} g$ all these moves are still projected along the local direction of steepest decent, i.e. $S$ is guaranteed to decrease at a sufficiently short step length. This modification to the fitting method ensures that $S$ decreases monotonically towards its local minimum value without any overshoot, ensuring that the approach to the best fit does not fail to converge.

### C. Assessing uncertainty in the best fit parameters

Once the data has been fitted the uncertainty in the fitting parameters about the best fit must be assessed. The change in $S$ ($\sigma_S$) associated with small changes in $\boldsymbol{\beta}$ ($\sigma_\beta$) is found by Taylor-expanding about the fitted minimum of $S(\boldsymbol{\beta})$.

$$S(\beta + \sigma_\beta) = S(\beta) + \boldsymbol{g}\sigma_\beta + \frac{1}{2}\mathbf{H}\sigma_\beta^2 + \cdots \quad (51)$$

$$\sigma_S = S(\beta + \sigma_\beta) - S(\beta) \quad (52)$$

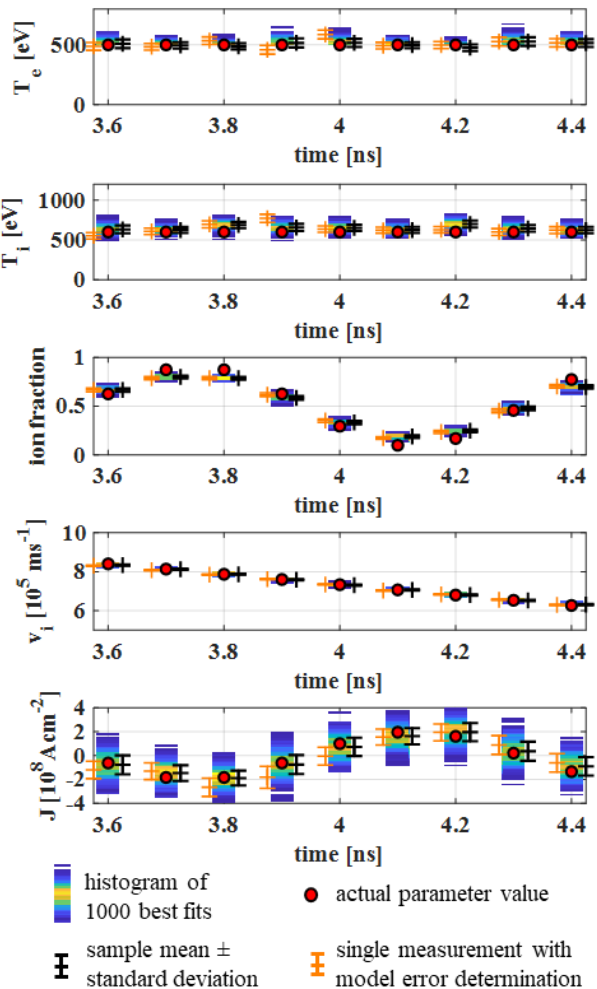

FIG. 5 Verification of the model for estimating the uncertainty in the best fit to the data. These plots compare the variation of underlying plasma parameters (red dots) to "measurements" simulated using the method outlined in section III.A. A histogram of the best fits for 1000 independent simulated measurements. The magnitude of sample standard deviation determined from this collection of fits (black) agrees well with the 1σ measurement error estimated using the analytical method outlined in III.B from a single simulated measurement (orange).

$$\sigma_S = \boldsymbol{g}\sigma_S(\boldsymbol{\beta}) + \frac{1}{2}\mathbf{H}\sigma_\beta^2 + \cdots \quad (53)$$

Since the gradient vector $\boldsymbol{g}$ approaches zero at the minimum value of $S$, the $\sigma_S$ associated with small changes in $\boldsymbol{\beta}$ can be reasonably approximated:

$$\sigma_S(\boldsymbol{\beta}) \approx \frac{1}{2}\mathbf{H}\sigma_\beta^2 \quad (54)$$

A $\sigma_S = 1$ is equivalent to an increase in the covariance of a single measurement by its entire expected error. This bounds the expected error of the fit. The covariance of the fitting parameters is therefore estimated by inverting this equation to give the covariance matrix for $\boldsymbol{\beta}$:

$$\sigma_\beta^2 \approx 2\mathbf{H}^{-1} \quad (55)$$

With the Hessian $\mathbf{H}$ calculated from the Jacobians at the best fit and the expected measurement covariance using equation (46). The square root of the diagonal elements of the resulting fitting parameter covariance matrix gives the errors in the fitting parameters:

$$\sigma_{\beta_i} = \sqrt{\sigma_\beta^2{}_{ii}} \quad (56)$$

For clarity, we note that $\sigma_{\beta_i}$ is the standard error, not the absolute error in the fit. The true value of the best fit has a 0.68 probability of lying within $\beta \pm \sigma_\beta$ and a 0.94 probability of lying within $2\sigma_\beta$.

### D. Verifying the estimate in the uncertainty of the best fit

The uncertainty that is determined using the method outlined in section IV.C can be compared to a sample variance of the fitting parameters generated by fitting many randomly generate synthetic streak images generated using the method outlined in section III.A. FIG. 5 shows an example of such a calculation, based on the synthetic data used to calculate the images in FIG. 3. The underlying, true plasma parameter used to generate the spectrum are plotted with a red dot. The 2D histogram shows the distribution of best fit values found using the method described in section IV.A to fit 1000 randomly generated images. The best fits found for each image where then used to calculate the sample mean and standard deviation of the fitting parameters, indicated by the black data point and associated error bars (this is offset from the histogram for readability). The orange data point shows an example of fit to a single individual synthetic image, with error bars determined using the method described in section IV.C.

Visual inspection of the data in FIG. 5 shows reasonably good agreement between the fitted values and the underlying true plasma parameters. The size of the error bars determined by the method described in IV.C are in excellent agreement with the sample standard deviations in the determination of the best fit, providing evidence that this method of estimating the fit error is valid.

Looking at the fits in more detail, we see a systematic failure of the analysis to capture the full amplitude of the ion fraction modulation. The analysis bins the data over 100ps, and therefore tends to blur out points of maximum amplitude for this signal, reducing the measured contrast somewhat. This observation highlights the compromises to the measurement accuracy that are introduced by temporal binning. Care must be taken to account for these effects in our interpretation of the data. The errors in the individual measurements of the plasma current are quite large for the signal amplitude of the data analyzed in the letter [8]. While the correlated modulation in the plasma current can be seen in the data, an accurate determination of the absolute magnitude of the current modulation cannot be determined at this signal level, with uncertainties in the best fit of ~50% of the peak amplitude.

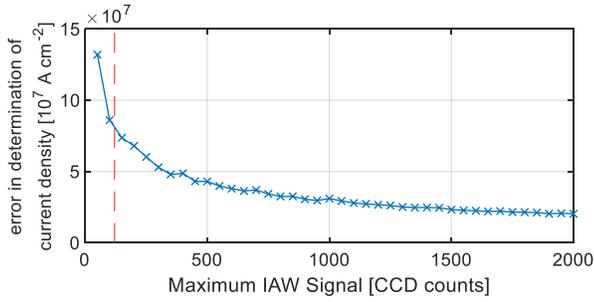

FIG. 6 Scaling of the error in the determination of the plasma current with IAW signal level. Red vertical line shows approximate max signal in data collected in our experiment.

Using the approach outlined in section IV.C it is simple to investigate how this measurement uncertainly scales with signal amplitude. Simulated profiles can be generated at a range of signal levels and fitted to extract an uncertainty. FIG. 6 plots the uncertainty as a function of varying maximum IAW signal. The red vertical line indicates the signal amplitude measured in the experiment. This plot indicated that a 4× improvement in the measurement precision could be achieved if the signal level could be improved by an order of magnitude. Measurement error increases rapidly at lower signal levels as

### E. Including other sources of error in the estimated covariance

Covariance matrices associated with different sources of measurement error can be added linearly. Accounting for additional measurement errors is therefore simple. As examples we consider the effects of the errors in the determination of the central wavelength of the spectrometer, and the error in the dispersion induced by drift in the magnification of the electron optics in the streak camera as a function of image position. The covariance due to the shift in the central wavelength can be estimated:

$$\boldsymbol{\sigma}_{shift} = f\left(x_i + \frac{\sigma_{x_0}}{2}, \boldsymbol{\beta}\right) - f\left(x_i - \frac{\sigma_{x_0}}{2}, \boldsymbol{\beta}\right)$$
$$\mathbf{K}_{shift} = \boldsymbol{\sigma}_{shift}\boldsymbol{\sigma}_{shift}^\dagger \quad (57)$$

While the covariance due to an error in the spectrometer magnification may be estimated:

$$\boldsymbol{\sigma}_{stretch} = f\left((x_i - x_0)\left(1 + \frac{\sigma_M}{2M}\right) + x_0, \boldsymbol{\beta}\right)$$
$$- f\left((x_i - x_0)\left(1 - \frac{\sigma_M}{2M}\right) + x_0, \boldsymbol{\beta}\right) \quad (58)$$
$$\mathbf{K}_{stretch} = \boldsymbol{\sigma}_{stretch}\boldsymbol{\sigma}_{stretch}^\dagger$$

The overall covariance used to determine the errors in the best fits is then simply:

$$\mathbf{K} = \mathbf{K}_{Noise} + \mathbf{K}_{Shift} + \mathbf{K}_{Stretch} \quad (59)$$

A typical example of these covariance matrices are shown in FIG. 7. The $\mathbf{K}_{Noise}$ matrix is close to diagonal and encodes correlation of pixels which are close together due to the coupling of the light from each photoelectron event to a small region of the CCD. The $\mathbf{K}_{Shift}$ matrix has large values along the diagonal close to the regions of maximum gradient in the signal, which correspond to large errors associated with a small change in the wavelength use in the calculation in these regions. There are regions of strong positive and negative covariance off the diagonal, encoding the expected correlation due to wavelength shift errors for regions with matching and opposite signal gradients respectively. The $\mathbf{K}_{Stretch}$ matrix is characterized by similar structures.

## V. APPLICATION OF FITTING METHOD TO EXISTING DATASET

In this section we provide a worked example applying the methods described in this paper to the data presented in FIG. 1.

### A. Determination of the noise parameters of the streak camera

To apply the techniques described in this paper it is first necessary to measure the noise parameters that characterize a given streak camera, the standard deviation of the read noise $\sigma_{RN}$, streak camera overall gain $G$ and noise factor $F^2$ and the point spread function (PSF) for a single photo-electron event (SPE) point spread function, $\sigma_{px}$. These characteristics can be ascertained by analyzing calibration images to extract the statistical properties of the camera.

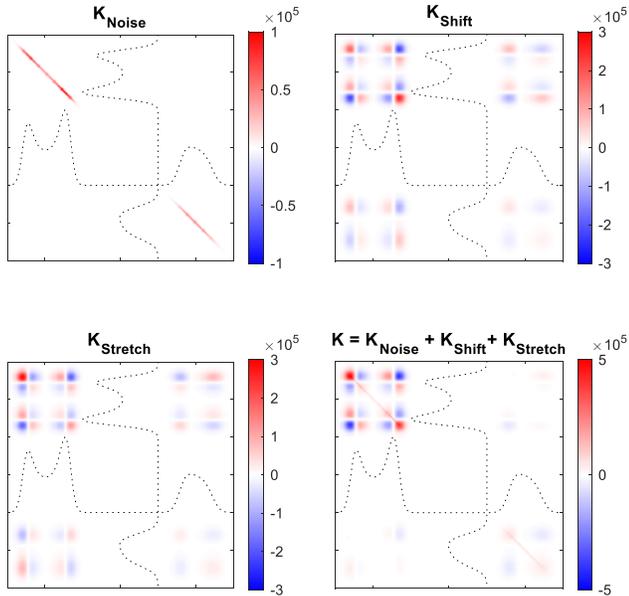

FIG. 7 Examples of noise, shift and stretch covariance matrices estimated for a calculated signal. The underlying calculate signal is plotted with dashed black lines both vertically and horizontally to aid in identifying the various of diagonal covariance terms.

The read noise for the streak camera CCD can be found by taking the standard deviation of pixels in the over-scan region of the image. The gain and noise factor for the streak camera can be assessed by constructing the pulse height distribution from a large set of sparse SPE calibration images. These are images taken at low signal levels, with typically < 100 SPEs per image. The method is described in detail in reference [18]. To summarize, images are first filtered to remove hot pixels, then events statistically above the noise floor (pixel values $> 5\sigma_{RN}$) of the images are detected via a threshold filter. These events are then sorted from highest to lowest. The total counts in the region within ±5 pixels of the detected event are summed. To prevent double counting this region is then overwritten with randomly generated noise of standard deviation $\sigma_{RN}$. Once all the detected events have been counted a histogram of event amplitude ("pulse height") is constructed. This histogram is then fitted using a model which accounts for null events ($D_0$), single photo-electron events ($D_1$) and double photo-electron events ($D_2$).

$$PHD(S) = \sum_{i=0}^{2} A_i \{D_i(s) * N_0(S)\}$$
$$D_0 = 1$$
$$D_1(S) = S \exp\left(-\frac{(S-\mu)^2}{2\sigma_{Added}^2}\right) \quad (60)$$
$$D_i(S) = D_{i-1} * D_1$$
$$N_0(S) = \exp\left(-\frac{S^2}{2\sigma_{RN}^2}\right)$$

Here $N_0(S)$ is the background distribution expected due to the read noise of the camera and $A_i$ are the amplitudes associated with each peak in the fit. The measured PHD is fitted with this function to extract values of model parameters $\mu$ and $\sigma_{Added}$. For optical streak cameras we can neglect generation of multiple electrons at the photocathode and therefore $G$ can be taken as the first moment of $D_1$.

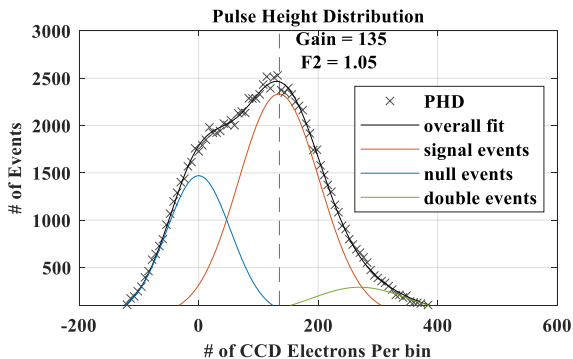

FIG. 8 Pulse height distribution for optical streak camera.

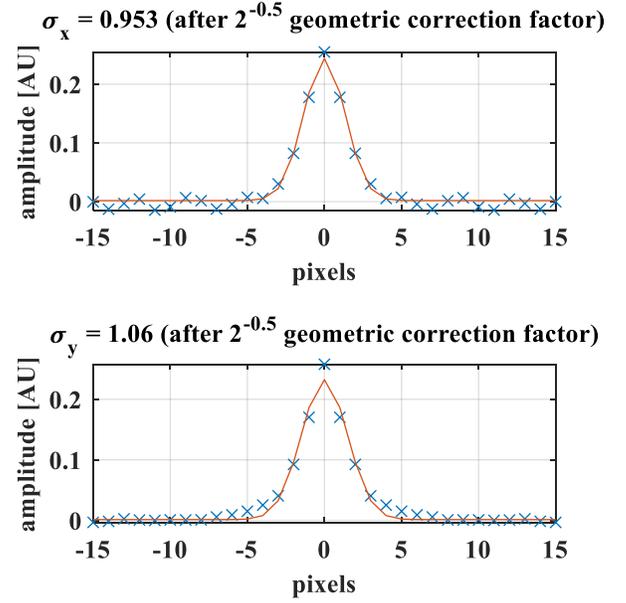

FIG. 9 Measuring the size of the single photo-electron event (SPE) point spread function (PSF) in both time and space direction.

$$G = \frac{\int_0^\infty S D_1 dS}{\int_0^\infty D_1 dS} \quad (61)$$

The Noise factor $F^2$ is then calculated:

$$F^2 = \sqrt{1 + \left(\frac{\sigma_{Added}}{G}\right)^2} \quad (62)$$

An example of a fit to such a PHD is illustrated in FIG. 8. The gain and noise factor extracted from this fit are $G = 135$ and $F^2 = 1.05$ respectively.

To determine the size of the PSF associated with each photoelectron event, summed profiles of the sparse SPE images are taken over 100 pixels widths of the images. These profiles are then used calculate a sample covariance matrix using equation (30). This matrix captures the covariance associated with three effects; the point spread function (PSF) of each SPE event, the read noise of the CCD sensor and any fluctuations in the effective brightness of the images in the overall set of sparse SPE images.

$$\mathbf{K} = \mathbf{K_{SPE}} + \mathbf{K_{RN}} + \mathbf{K_F} \quad (63)$$

The covariance associated with the read noise ($\mathbf{K_{RN}}$) can be determined by carrying out the same analysis on a set background images, where no signal is present (i.e. streak electron optics are off). The brightness fluctuations are approximately uniform over the whole image and therefore $\mathbf{K_F}$ can be approximated as uniform, with an amplitude estimated by taken the mean of the regions of $\mathbf{K}$ far from the diagonal.

Once these unwanted components have been subtracted the PSF profile can be constructed by taking the mean of each diagonal of the covariance matrix. The resulting profile can then be fitted to find the width of the SPE. This whole analysis can be carried out in both time and space directions of the streak camera to independently measure the width of the PSF in each direction. An example of this final fit is provided in FIG. 9. The shape of the PSF is modeled reasonably accurately by a gaussian, but this model could be improved if desired. The fitted width must be corrected by a geometric factor of $2^{-0.5}$ to get the true standard deviation of the point spread function. Taking the mean of the two measurements gives a $\sigma_{px} = 1.0$ px.

**B. Example fitting of the data**

The approach to fitting outlined in this paper was applied to the fitting of the experimental data presented in FIG. 1. The results are presented in FIG. 10. The electron density in this plot is measured by fitting the electron plasmas wave feature of the TS spectrum. The error bars shown in the figure represent solely the errors associated with fitting in the presence of the streak camera measurement noise. If, additionally, we allow for an error in the central wavelength of 0.1 nm and uncertainty in the magnification of streak camera of 0.02, then the errors in the determination of the flow velocity roughly double. This is simply because the flow velocity is diagnosed directly from the magnitude of the doppler shift, which is shifted by these uncertainties. Errors in the other parameters are largely unaffected. The errors determined by this method are largely in line with the magnitude of the errors reported in our earlier letter.

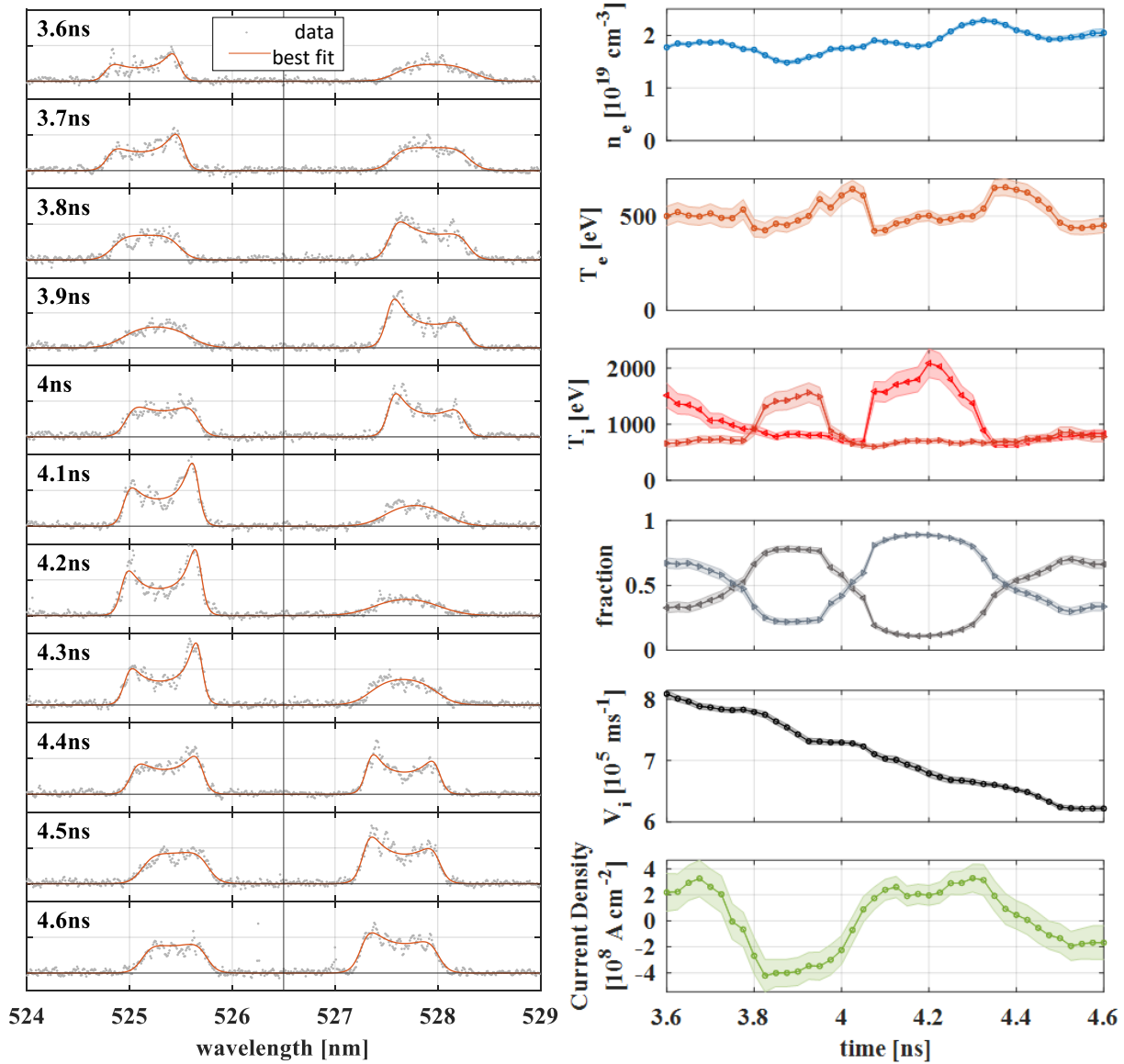

FIG. 10 Experimental data plotted in FIG. 1 is fitted using the method outlined in this paper. The variations in the fitting parameters are illustrated along with the associated one-σ error bars as shaded regions. The plots of $T_i$ and fraction show the variations for the red-shifted (◄) and blue-shifted (►) features independently.

It is important to reiterate that the uncertainties determined using the method outlined in this paper are uncertainties only in the best fit of a specific model to the measured data. Making accurate measurements of plasma dynamics in HED experiments using Thomson scattering requires a holistic approach, taking account of many interconnected contributing factors. The methods described here cannot account for errors associated with using an incorrect or inadequate model of the plasma behavior. Careful theoretical analysis of the plasma dynamics and scattering physics are required to justify the models used to calculate the expected scattered spectra. It is only with this strong theoretical foundation that the methods described can be used with confidence. This caveat also extends also to situations where the signal is contaminated by some unwanted background signal, arising due to plasma self-emission or through some laser plasma instability.

## VI. CONCLUSIONS

This paper presents a mathematical framework for understanding and modeling measurement noise introduced by streak cameras. A method of accounting for this noise in the fitting process has been presented that allows the calculation of accurate uncertainties in the best fit of the model to this inherently noisy experimental data. The models presented in this paper have been verified using random number generators to simulate the streak camera noise generation process and examine the resulting statistics.

The model allows an expected covariance matrix (eqn. (38)) to be calculated directly from the proposed fit to the data. This covariance matrix is used to weight the fitting algorithm and is updated iteratively as the best fit is approached. Once found, the errors in the determination of the best fit can be extracted from the final covariance matrix to provide measurement errors for the analysis.

The methods developed here can also be used to estimate how the errors associated with specific measurements of the underlying plasma parameters scale with the amplitude of the measured signal. This is important tool for experimental design and for quantitatively defining the signal requirements for future Thomson scattering diagnostics.

The detailed presentation of the origins and effects of streak camera measurement noise presented here should provide a firm foundation on which to build confidence in future measurements where signal to noise is low and the fits to the data are therefore open to potential criticism. It is hoped that the thorough mathematical description developed here may be wielded as a tool to justify the validity and significance of otherwise seemingly tenuous measurements such as the current density measurement presented in our previous paper [8].

## VII. ACKNOWLEDGMENTS

This work was performed under the auspices of the U.S. Department of Energy by Lawrence Livermore National Laboratory under Contract DE-AC52-07NA27344.

This document was prepared as an account of work sponsored by an agency of the United States government. Neither the United States government nor Lawrence Livermore National Security, LLC, nor any of their employees makes any warranty, expressed or implied, or assumes any legal liability or responsibility for the accuracy, completeness, or usefulness of any information, apparatus, product, or process disclosed, or represents that its use would not infringe privately owned rights. Reference herein to any specific commercial product, process, or service by trade name, trademark, manufacturer, or otherwise does not necessarily constitute or imply its endorsement, recommendation, or favoring by the United States government or Lawrence Livermore National Security, LLC. The views and opinions of authors expressed herein do not necessarily state or reflect those of the United States government or Lawrence Livermore National Security, LLC, and shall not be used for advertising or product endorsement purposes.

## VIII. DATA AVAILABILITY STATEMENT

The data that support the findings of this study are available from the corresponding author upon reasonable request.

## IX. REFERENCES AND FOOTNOTES

[1] J. Sheffield, D. Froula, S. H. Glenzer, N. C. Luhmann, and Jr., *Plasma Scattering of Electromagnetic Radiation: Theory and Measurement Techniques*, 2nd ed., Vol. 2010 (Academic Press, 2010).

[2] S. H. Glenzer, C. A. Back, L. J. Suter, M. A. Blain, O. L. Landen, J. D. Lindl, B. J. MacGowan, G. F. Stone, R. E. Turner, and B. H. Wilde, *Thomson Scattering from Inertial-Confinement-Fusion Hohlraum Plasmas*, Physical Review Letters **79**, 1277 (1997).

[3] S. H. Glenzer, W. Rozmus, B. J. MacGowan, K. G. Estabrook, J. D. De Groot, G. B. Zimmerman, H. A. Baldis, J. A. Harte, R. W. Lee, E. A. Williams, and B. G. Wilson, *Thomson Scattering from High-Z Laser-Produced Plasmas*, Physical Review Letters **82**, 97 (1999).

[4] D. Froula, J. Ross, B. Pollock, P. Davis, A. James, L. Divol, M. Edwards, A. Offenberger, D. Price, R. Town, G. Tynan, and S. Glenzer, *Quenching of the Nonlocal Electron Heat Transport by Large External Magnetic Fields in a Laser-Produced Plasma Measured with Imaging Thomson Scattering*, Physical Review Letters **98**, 135001 (2007).

[5] J. S. Ross, H.-S. Park, R. Berger, L. Divol, N. L. Kugland, W. Rozmus, D. Ryutov, and S. H. Glenzer, *Collisionless Coupling of Ion and Electron Temperatures in Counterstreaming Plasma Flows*, Physical Review Letters **110**, 145005 (2013).

[6] G. F. Swadling, S. V. Lebedev, A. J. Harvey-Thompson, W. Rozmus, G. C. Burdiak, L. Suttle, S. Patankar, R. A. Smith, M. Bennett, G. N. Hall, F. Suzuki-Vidal, and J. Yuan, *Interpenetration, Deflection, and Stagnation of Cylindrically Convergent Magnetized Supersonic Tungsten Plasma Flows*, Physical Review Letters **113**, 035003 (2014).

[7] C. Neuville, V. Tassin, D. Pesme, M.-C. Monteil, P.-E. Masson-Laborde, C. Baccou, P. Fremerye, F. Philippe, P. Seytor, D. Teychenné, W. Seka, J. Katz, R. Bahr, and S. Depierreux, *Experimental Evidence of the Collective Brillouin Scattering of Multiple Laser Beams Sharing Acoustic Waves*, Physical Review Letters **116**, (2016).

[8] G. F. Swadling, C. Bruulsema, F. Fiuza, D. P. Higginson, C. M. Huntington, H.-S. Park, B. B. Pollock, W. Rozmus, H. G. Rinderknecht, J. Katz, A. Birkel, and J. S. Ross, *Measurement of Kinetic-Scale Current Filamentation Dynamics and Associated Magnetic Fields in Interpenetrating Plasmas*, Physical Review Letters **124**, 215001 (2020).

[9] A. M. Hansen, D. Turnbull, J. Katz, and D. H. Froula, *Mitigation of Self-Focusing in Thomson Scattering Experiments*, Physics of Plasmas **26**, 103110 (2019).

[10] W. A. Farmer, C. Bruulsema, G. F. Swadling, M. W. Sherlock, M. D. Rosen, W. Rozmus, D. H. Edgell, J. Katz, B. B. Pollock, and J. S. Ross, *Validation of Heat Transport Modeling Using Directly Driven Beryllium Spheres*, Physics of Plasmas **27**, 082701 (2020).

[11] C. Bruulsema, W. Rozmus, G. F. Swadling, S. Glenzer, H. S. Park, J. S. Ross, and F. Fiuza, *On the Local Measurement of Electric Currents and Magnetic Fields Using Thomson Scattering in Weibel-Unstable Plasmas*, Physics of Plasmas **27**, 052104 (2020).

[12] A. L. Milder, J. Katz, R. Boni, J. P. Palastro, M. Sherlock, W. Rozmus, and D. H. Froula, *Measurements of Non-Maxwellian Electron Distribution Functions and Their Effect on Laser Heating*, Phys. Rev. Lett. **127**, 015001 (2021).

[13] C. T. Silbernagel, P. Torres III, and D. H. Kalantar, *A Method for Analyzing High-Resolution Time-Domain Streak Camera Calibration Data*, in edited by F. T. Luk (Denver, CO, 2004), p. 435.

[14] D. E. Evans and J. Katzenstein, *Laser Light Scattering in Laboratory Plasmas*, Reports on Progress in Physics **32**, 305 (1969).

[15] P. A. Jaanimagi, R. Boni, D. Butler, S. Ghosh, W. R. Donaldson, and R. L. Keck, *The Streak Camera Development Program at LLE*, in edited by D. L. Paisley, S. Kleinfelder, D. R. Snyder, and B. J. Thompson (Alexandria, VA, 2005), p. 408.

[16] Swadling, George, *LLNL/Thomson-Scattering-Cross-Section-Calculator* (Lawrence Livermore National Laboratory (LLNL), Livermore, CA (United States), 2021).

[17] A. Visco, R. P. Drake, D. H. Froula, S. H. Glenzer, and B. B. Pollock, *Temporal Dispersion of a Spectrometer*, Review of Scientific Instruments **79**, (2008).

[18] S. Ghosh, R. Boni, and P. A. Jaanimagi, *Optical and X-Ray Streak Camera Gain Measurements*, Review of Scientific Instruments **75**, 3956 (2004).

[19] R. A. Lerche, J. W. McDonald, R. L. Griffith, G. V. de Dios, D. S. Andrews, A. W. Huey, P. M. Bell, O. L. Landen, P. A. Jaanimagi, and R. Boni, *Preliminary Performance Measurements for a Streak Camera with a Large-Format Direct-Coupled Charge-Coupled Device Readout*, Review of Scientific Instruments **75**, 4042 (2004).

[20] A. Bjorck, *Numerical Methods for Least Squares Problems* (SIAM, 1996).

[21] J. Nocedal and S. Wright, *Numerical Optimization* (Springer Science & Business Media, 2000).